# Optically Pumped Floquet States of Magnetization in Ferromagnets

I.V. Savochkin[1, 2], A.K. Zvezdin[2, 3], V.I. Belotelov[1, 2, a]

[1]*Lomonosov Moscow State University, 119991 Moscow, Russia*
[2]*Russian Quantum Center, Skolkovo, 143025 Moscow Region, Russia*
[3]*Prokhorov General Physics Institute, Russian Academy of Sciences, 119991 Moscow, Russia*

[a]Corresponding author: belotelov@physics.msu.ru

**Abstract.** Floquet states have been subject of great research interest since Zel'dovich's pioneering work on the quasienergy of a quantum system subject to a temporally periodic action. Nowadays periodic modulation of the system Hamiltonian is mostly achieved by microwaves leading to novel exciting phenomena in condensed matter physics: Floquet topological insulators, chiral edge states etc. On the other hand, nonthermal optical control of magnetization at picosecond time scales is currently a highly appealing research topic for potential applications in magnetic data storage. Here we combine these two concepts to investigate Floquet states in the system of exchange-coupled spins in a ferromagnet. We periodically perturb the magnetization of an iron-garnet film by a train of circularly-polarized femtosecond laser pulses hitting the sample at 1 GHz repetition rate and monitor the magnetization dynamics behaving like a Floquet state. An external magnetic field allows tuning of the Floquet states leading to a pronounced increase of the precession amplitude by one order of magnitude at the center of the Brillouin zone, i.e. when the precession frequency is a multiple of the laser pulse repetition rate. Floquet states might potentially allow for parametric generation of magnetic oscillations. The observed phenomena expand the capabilities of coherent ultrafast optical control of magnetization and pave a way for their application in quantum computation or data processing.

## INTRODUCTION

Currently, temporally periodic perturbations of a quantum system that lead to Floquet states are of prime interest in frontier research areas such as topological insulators in graphene, or Bose-Einstein condensates with cold atoms [1-6]. New surprising phenomena with tantalizing properties appear: periodic perturbation of a semiconductor quantum well by microwaves might induce a Floquet topological state [1-4], or microwave induced dynamics of a two-level artificial atom may enable elaborated quantum control [5].

In this work we investigate Floquet states in experiments on optical excitation of magnetization precession in magnetically ordered media. Among different mechanisms responsible for optical pumping the magnetization dynamics in ferromagnets [7-25], the inverse Faraday effect is of particular importance [18-25]. It can be observed through the impact of circularly polarized light on the medium magnetization, as if it would be induced by an effective magnetic field $\mathbf{H}_F \sim [\mathbf{E}\times\mathbf{E}^*]$, where $\mathbf{E}$ is the electric field of the light wave [26]. The field $\mathbf{H}_F$ is directed along the light wave vector. In ferromagnets it originates microscopically from stimulated Raman scattering on magnons.

The inverse Faraday effect is observed in pump-probe experiments [18-25]: The magnetization dynamics excited by the pump is observed by a low intensity probe pulse sent onto the sample delayed relative to the pump. The high temporal resolution of the pump-probe technique allowed identification of novel dynamic magnetic properties of ferromagnets. Moreover, the authors of Ref. 23 managed to optically excite magnetostatic spin waves in iron garnets and later their tunability by beam diameter was demonstrated [27, 28].

Recently, influence of the periodic pumping has been demonstrated [29, 30]. It was shown that a train of optical pump pulses can drive the magnetization within the illuminated area thereby exciting spin waves and their amplitude can be significantly increased if the magnetization oscillation frequency synchronized with the laser pulses. Moreover, the periodic pumping additionally brings several important features for spin waves propagation: it significantly increases their propagation distance and provides also the possibility to tune directionality of propagation [29], and also tune their wavelength by sweeping the external magnetic field or by slight variation of the pulse repetition rate [30].

Spin waves amplitude generated in that experiments corresponded to the precession angle of the order of 10 µrad which was due to the relatively small effective magnetic field of the inverse Faraday effect, ~ 10 Oe. However, influencing the sample with a train of pulses inducing much larger periodic magnetic field could lead to some novel phenomena. Here we consider this issue theoretically and show a possibility of excitation of Floquet states of the magnetization in iron-garnet films by a sequence of laser pump pulses. By varying an external magnetic field it is possible to modify and control the Floquet states.

## FLOQUET STATES OF MAGNETIZATION DYNAMICS

The excitation of magnetization dynamics by optical pulses generally leads to a spatial and temporal pattern of the magnetization $\mathbf{M}(\mathbf{r}, t)$ that is a superposition of different modes $\mathbf{M}_q(\mathbf{r}, t)$. The magnetization dynamics of these modes is given by [31]

$$i\hbar \frac{d\mathbf{M}_q}{dt} = -\left[\hat{H}, \mathbf{M}_q\right], \qquad (1)$$

where $[A, B] = AB - BA$ is the commutator of operators $A$ and $B$, and $\hat{H}$ is the Hamiltonian of the spin system. The Hamiltonian has contributions from the interaction of the magnetization with $\mathbf{H}_F$. Optical excitation with a period $T$ leads also to $\hat{H}(t)$ becoming periodic: $\hat{H}(t + T) = \hat{H}(t)$. In accordance to Floquet theorem, the eigensolutions can be represented as $\mathbf{M}_q(t + T) = e^{-i\Omega T} \mathbf{M}_q(t)$, where $\Omega$ is called quasifrequency in analogy to the quasimomentum of the Bloch functions in real space [32].

Let us consider an experiment where an iron-garnet film placed into external magnetic field $\mathbf{H}$ shooted by a sequence of circularly polarized laser pulses (Fig. 1). Due to the inverse Faraday effect such pulses will influence on magnetization like a pulses of magnetic field $\mathbf{H}_F$ and will excite periodic magnetization dynamics. Magnetization vector $\mathbf{M}$ starts to precess around its equilibrium position and amplitude of this precession could be described by angle $\theta$.

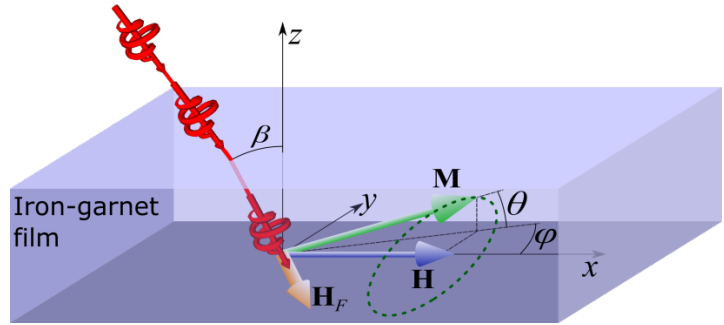

***FIGURE 1.*** *Excitation of the magnetization precession in an iron-garnet film by a train of optical pulses.*

Without illumination the sample is uniformly magnetized along the effective magnetic field $\mathbf{H}_e = \mathbf{H} + \mathbf{H}_a$, with the magnetocrystalline anisotropy field $H_a = 4\pi M_s - 2K_U/M_s$, where $M_s$ – saturation magnetization and $K_U$ – uniaxial anisotropy constant. When a laser pulse propagates through the film its impact on the magnetization can be described in terms of $\mathbf{H}_F$. If $\mathbf{H}$ is in the sample plane, Eq. (1) gives [29]:

$$\frac{\partial^2 \theta}{\partial t^2} + \frac{2}{\tau}\frac{\partial \theta}{\partial t} + \left[\omega_0^2 - V(t)\right]\theta = \gamma^2 H H_F(t)\cos\beta, \qquad (2)$$

where $\gamma$ is the gyromagnetic ratio, $\tau = 2(\alpha(\omega_H+\omega_e))^{-1}$ is the decay time, $\alpha$ is Gilbert damping constant, $\omega_0 = \sqrt{\omega_H \omega_e}$, $\omega_H = \gamma H$, $\omega_e = \gamma(H + H_a)$, $V(t) = -(\omega_H+\omega_e)\gamma H_F(t)\sin\beta$ and $\beta$ is the angle of the light wave vector in the magnetic film. In our experiment the optical pulse duration $\Delta t \ll T, \omega_0^{-1}$, justifying the following representation: $H_F(t) = h\Delta t \sum_{m=-\infty}^{+\infty} \delta(t - mT)$, where $h$ is the amplitude of $H_F$, and $m$ is an integer.

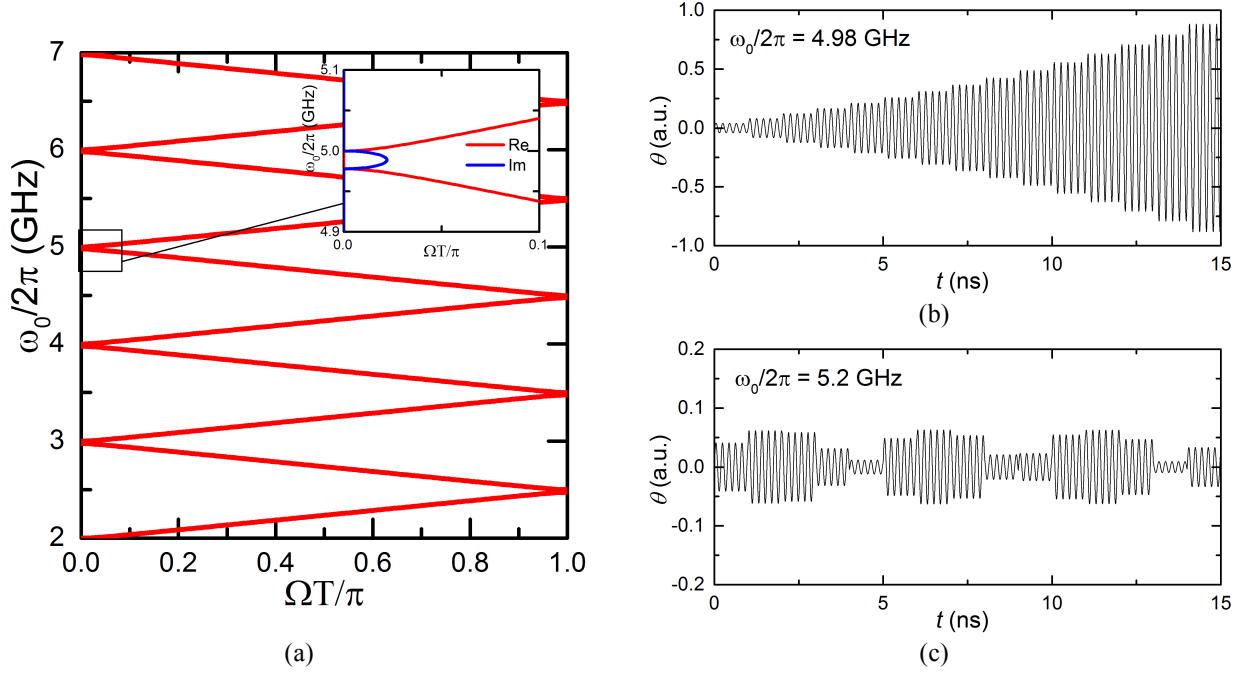

**FIGURE 2.** Optically induced Floquet states for the magnetization mode. (a) Calculated spectrum of the Floquet states: Re($\Omega$) (red curves) and Im($\Omega$) (blue curve). (b, c) Parametrically excited oscillations at $\omega_0/2\pi = 4.98$ GHz (b) and $\omega_0/2\pi = 5.2$ GHz (c). Time $t$ is counted starting from the first pump pulse. For calculations we take $H_a = 1800$ Oe, $h = 1000$ Oe, $\Delta t = 5$ ps, $\beta = 45°$.

In case of negligible magnetic losses $\tau \to \infty$ and tangent pump pulse incidence $\beta \to \pi/2$ Eq. (2) becomes isomorphic to the Schrödinger equation with a periodic potential $V(t) = V_0 \Delta t \sum_{m=-\infty}^{+\infty} \delta(t - mT)$, and the problem is reduced to the Kronig-Penney model with eigenenergies $\omega_0^2$:

$$-\frac{\partial^2 \theta}{\partial t^2} + V(t)\theta = \omega_0^2 \theta. \quad (3)$$

The solution of Eq. (3) can be written as $\theta(t) = \vartheta(t) e^{i\Omega t}$ with the periodic Floquet amplitude $\vartheta(t+T) = \vartheta(t)$. Assuming $\Delta t \to 0$ and $V_0 \to \infty$ (but $V_0 \Delta t$ remains constant) we could derive the following transcendental equation describing $\omega_0(\Omega)$ dependence:

$$\cos \Omega T = \frac{1}{2} V_0 \Delta t \frac{\sin \omega_0 T}{\omega_0} + \cos \omega_0 T, \quad (4)$$

where $V_0 = -\gamma h(\omega_H + \omega_e)$ – amplitude of periodic potential. Since $\omega_0$ depends on the external magnetic field $H$, the Floquet states are tunable by this field. Their spectrum, $\omega_0 = \omega_0(\Omega)$, in the first Brillouin zone is characteristic for "periodic potentials" (Fig. 2a) with bandgaps whose size is proportional to $V_0 \Delta t$ and could be calculated as:

$$\Delta \omega = \frac{V_0 \Delta t}{\pi N_g}, \quad (5)$$

where $N_g$ is bandgap number. Our theoretical analysis predicts that for $\omega_0$ within the bandgap parametric generation of magnetization oscillations will take place (Fig. 2b). Then the quasifrequency $\Omega$ becomes purely imaginary (inset of Fig. 2a) and the oscillation amplitude increases exponentially as long as the non-linear limitation does not come into play. And outside of bandgaps the magnetization oscillations have envelope with frequency $\Omega$ (Fig. 2c).

In experiments described in [29, 30] the optical pumping was realized in thin bismuth iron-garnet films [33] excited magnetic field *h* was about 10 Oe, so the bandgap was very small and due to the damping no Floquet states were observed. But the pump incidence angle was 17°, and main effect of amplitude increasing was from the right part of Eq. (3) $\gamma^2 H H_F(t) \cos\beta$, acting like an external force.

## CONCLUSION

We have shown that for such periodic excitation a peculiar magnetization oscillation regime corresponding to Floquet states could be achieved. The Floquet states are tunable via an external magnetic field and manifest themselves in a quasi-periodic dependence of amplitude and phase on magnetic field. At a resonance, the precession amplitude is considerably increased. In terms of the Floquet state spectrum this enhancement occurs for an oscillation quasifrequency at the center of the Brillouin zone. In other words, the amplitude resonances take place when the oscillations induced by subsequent pulses become synchronized to each other. Another picturesque consequence of the Floquet states might be parametric generation. For demonstration this requires, however, much higher pump fluencies and longer pulse durations than were available in the current experiment.

Generally, optical excitation of magnetic oscillations provides control of the magnetization distribution at submicron space and sub-terahertz time scales which is of prime importance for quantum information processing based on spin waves [34]. Tunable Floquet states of magnetization oscillations might significantly broaden the functionality of this approach. The next step forward in this paradigm will be implementation of plasmonic structures to further enhance the magnetization precession by concentrating the optical fields in a nanometer thick magnetic film. Indeed, while the huge plasmonics-mediated increase of the direct magneto-optical effects was demonstrated recently [35], the plasmonic boost of the inverse magneto-optical effects is still waiting for its practical implementation [9].

## ACKNOWLEDGEMENTS

This work was financially supported by the Russian Science Foundation (Grant No. 17-72-20260).

## REFERENCES

1. Lindner, N. H. et al. Floquet topological insulator in semiconductor quantum wells. Nat. Phys. **7**, 490-495 (2011).
2. Rechtsman, M. C. et al. Photonic Floquet topological insulators. Nature **496**, 196-200 (2013)
3. Wang, Y. H. et al. Observation of Floquet-Bloch States on the Surface of a Topological Insulator Science **342**, 453-457 (2013).
4. Sentef, M. A. et al. Theory of Floquet band formation and local pseudospin textures in pump-probe photoemission of graphene. Nat. Commun. **6**, 7047 (2015).
5. Deng, C. et al. Observation of Floquet States in a Strongly Driven Artificial Atom. Phys. Rev. Lett. **115**, 133601 (2015).
6. Jotzu, G. et al. Experimental realization of the topological Haldane model with ultracold fermions. Nature **515**, 237-240 (2014).
7. Beaurepaire, E. et al. Ultrafast Spin Dynamics in Ferromagnetic Nickel. Phys. Rev. Lett. **76**, 4250-4253 (1996).
8. Kalashnikova, A. M. et al. Impulsive excitation of coherent magnons and phonons by subpicosecond laser pulses in the weak ferromagnet FeBO3. Phys. Rev. B **78**, 104301 (2008).
9. Belotelov, V. I. and Zvezdin, A. K. Inverse transverse magneto-optical Kerr effect. Phys. Rev. B **86**, 155133 (2012).
10. Bigot, J.-Y. et al. Coherent ultrafast magnetism induced by femtosecond laser pulses. Nat. Phys. **7**, 515-520 (2009).
11. Kimel, A. V. et al. Laser-induced ultrafast spin reorientation in the antiferromagnet TmFeO3. Nature **429**, 850-853 (2004).
12. Koene, B. et al. Excitation of magnetic precession in bismuth iron garnet via a polarization-independent impulsive photomagnetic effect. Phys. Rev. B **91**, 184415 (2015).
13. van Kampen, M. et al. All-Optical Probe of Coherent Spin Waves. Phys. Rev. Lett. **88**, 227201 (2002)

14. Stamm, C. et al. Femtosecond modification of electron localization and transfer of angular momentum in nickel. Nat. Mater. **6**, 740-743 (2007).
15. Yoshimine, I. et al. Phase-controllable spin wave generation in iron garnet by linearly polarized light pulses. J. Appl. Phys. **116**, 043907 (2014).
16. Atoneche, F. et al. Large ultrafast photoinduced magnetic anisotropy in a cobalt-substituted yttrium iron garnet. Phys. Rev. B **81**, 214440 (2010)
17. Bossini, D. et al. Magnetoplasmonics and Femtosecond Optomagnetism at the Nanoscale. ACS Photonics **3**, 1385-1400 (2016).
18. Kimel. A. V. et al. Ultrafast non-thermal control of magnetization by instantaneous photomagnetic pulses. Nature **435**, 655-657 (2005).
19. Stanciu, C. D. et al. All-Optical Magnetic Recording with Circularly Polarized Light. Phys. Rev. Lett. **99**, 047601 (2007).
20. Hansteen, F. et al. Nonthermal ultrafast optical control of the magnetization in garnet films. Phys. Rev. B **73**, 014421 (2006).
21. Reid, A. H. M. et al. Optical Excitation of a Forbidden Magnetic Resonance Mode in a Doped Lutetium-Iron-Garnet Film via the Inverse Faraday Effect. Phys. Rev. Lett. **105**, 107402 (2010).
22. Parchenko, S. et al. Wide frequencies range of spin excitations in a rare-earth Bi-doped iron garnet with a giant Faraday rotation. Appl. Phys. Lett. **103**, 172402 (2013).
23. Satoh, T. et al. Directional control of spin-wave emission by spatially shaped light. Nat. Photonics. **6**, 662-666 (2012).
24. Iida, R. et al. Spectral dependence of photoinduced spin precession in DyFeO3. Phys. Rev. B **84**, 064402 (2011).
25. Mikhaylovskiy R. V. et. al. Ultrafast inverse Faraday effect in a paramagnetic terbium gallium garnet crystal. Phys. Rev. B **86**, 100405 (2012).
26. L. P. Pitaevskii, Zh. Eksp. Teor. Fiz. **39**, 1450 (1960) [Sov. Phys. JETP **73**, 672 (1991)].
27. Chernov A.I. et al. Optical excitation of spin waves in epitaxial iron garnet films: MSSW vs BVMSW. Opt. Lett. **42**, 279-282 (2017).
28. Chernov A.I. et al. Local probing of magnetic films by optical excitation of magnetostatic waves. Phys. Solid State **58**, 1128-1134 (2016).
29. Jäckl, M. et al. Magnon Accumulation by Clocked Laser Excitation as Source of Long-Range Spin Waves in Transparent Magnetic Films. Phys. Rev. X **7**, 21009 (2017).
30. Savochkin, I. V. et al. Generation of spin waves by a train of fs-laser pulses: a novel approach for tuning magnon wavelength. Scientific reports **7**, 5668 (2017).
31. Akhiezer, A. I. et al., Spin Waves (North-Holland, Amsterdam, 1968).
32. Zel'dovich, Ya. B. Scattering and emission of a quantum system in a strong electromagnetic wave, Sov. Phys. Usp. **16**, 427 (1973).
33. Vasiliev, M. et al. RF magnetron sputtered $(BiDy)_3(FeGa)_5O_{12}:Bi_2O_3$ composite garnet-oxide materials possessing record magneto-optic quality in the visible spectral region. Opt. Express **17**, 19519-19535 (2009).
34. Wu, H. et al. Storage of Multiple Coherent Microwave Excitations in an Electron Spin Ensemble. Phys. Rev. Lett. **105**, 140503 (2010).
35. Belotelov, V. I., Zvezdin, A. K. Magnetooptics and extraordinary transmission of the perforated metallic films magnetized in polar geometry. J. Magn. Magn. Mater. **300**, 260-263 (2006).